\documentclass[journal,draftclsnofoot,onecolumn,12pt]{IEEEtran}
\usepackage[cmex10]{amsmath}
\usepackage{setspace}
\usepackage{amsfonts}
\usepackage{amssymb}
\usepackage{subcaption}

\usepackage{algorithm}
\usepackage{algpseudocode}
\usepackage{pifont}

\usepackage{cite}      
\usepackage{graphicx}  

\usepackage{citesort}

\usepackage{setspace}
\begin{document}
\linespread{0.95}

\title{Spectrally and Energy Efficient OFDM (SEE-OFDM) for Intensity Modulated Optical Wireless Systems}

\author{Emily~Lam,~\IEEEmembership{}
        Sarah~Kate~Wilson,~\IEEEmembership{}
        Hany~Elgala,~\IEEEmembership{}
				and~Thomas~D.~C.~Little.~\IEEEmembership{}
\thanks{E. Lam and T.D.C. Little are with the Department of Electrical and Computer Engineering, Boston University, Boston, MA, 02215, USA (e-mail: \{emilylam,tdcl\}@bu.edu).}
\thanks{S.K. Wilson is with the Electrical Engineering Department, Santa Clara University, Santa Clara, CA, 95053, USA (e-mail: skwilson@scu.edu).}
\thanks{H. Elgala is with the Computer Engineering Department, University at Albany, Albany, NY, 12222, USA (e-mail: helgala@albany.edu).}%
}

\maketitle

\begin{abstract}
Spectrally and energy efficient orthogonal frequency division multiplexing (SEE-OFDM) is an optical OFDM technique based on combining multiple asymmetrically clipped optical OFDM (ACO-OFDM) signals into one OFDM signal. By summing different components together, SEE-OFDM can achieve the same spectral efficiency as DC-biased optical OFDM (DCO-OFDM) without an energy-inefficient DC-bias. This paper introduces multiple methods for decoding a SEE-OFDM symbol and shows that an iterative decoder with hard decisions gives the best performance. Being a multi-component format, different energy allocation amongst the different components of SEE-OFDM is possible. However, equal energy allocation performs 1.5 dB better than unequal energy allocation. A hard-decision, iterative subtraction receiver can further increase performance by another 1.5 dB over soft-decision subtraction and reconstruction receivers. SEE-OFDM consistently performs 3 dB or better and with higher spectral efficiency than ACO-OFDM at the same bit-error-rate (BER). Comparing other combination methods at the same BER, SEE-OFDM performs up to 3 dB better than hybrid asymmetrically clipped optical (OFDM) (HACO-OFDM) and up to 1.5 dB better than asymmetrically and symmetrically clipped optical OFDM (ASCO-OFDM) and enhanced unipolar OFDM (eU-OFDM) when using hard decisions at the receiver. Additionally, SEE-OFDM has the best peak-to-average-power rate (PAPR) as compared to the other combination OFDM formats and ACO-OFDM, which makes it excellent for any range limited optical source, such as laser diodes and light-emitting diodes (LEDs). In summary, SEE-OFDM is shown to have excellent properties to glean additional capacity from an intensity modulation and direct detection (IM/DD) optical wireless communications system.
\end{abstract}


\IEEEpeerreviewmaketitle

\section{Introduction}\label{introduction}
\IEEEPARstart{O}{ptical} wireless communications (OWC) systems deal
primarily with the IR, Visible, and UV bands of the electromagnetic
spectrum and can employ a variety of the properties of light to gain
spectral efficiency. These include coherence, polarization, and
orbital angular momentum. However, these techniques are not practical
for many low-cost optical sources including LEDs or laser diodes. The
alternative approach is to use use incoherent intensity modulation and
direct detection (IM/DD) of the optical signal. The goal with this
approach is to make available new capacity by using low-cost optical
devices that can operate in the visible spectrum. By enabling simple
and efficient modulation at this operating point we can begin to
address the challenge of increased data consumption in wireless data
systems that is affecting the world today.  \cite{emh09}.

For IM/DD OWC systems, direct quadrature modulation is not possible. This is due to the fact that signals are modulated and recovered only by varying and detecting the instantaneous intensity of the light \cite{kb97}. As intensity cannot be negative, transmission of negative values are not possible. These two criteria constrict OWC for IM/DD systems to real and positive valued signals. There exists unipolar real modulation schemes, such as on-off keying (OOK) and pulse-position modulation (PPM), for use in IM/DD systems \cite{gel13}. However, these existing schemes are single-carrier schemes, which require complex equalization at the receiver compared to multi-carrier schemes such as orthogonal frequency division multiplexing (OFDM). Prior methods for modulating OFDM in an IM/DD system, namely the foundational DC-biased optical OFDM (DCO-OFDM) and asymmetrically clipped optical OFDM (ACO-OFDM), have drawbacks with respect to energy and spectral efficiency \cite{gpr05,al06}.

Recent research efforts have focused on optical OFDM techniques that optimize spectral and energy efficiency \cite{da13,rk14,wb15,th14}. Most of these techniques are combination techniques, that is techniques that sum together separate OFDM signals based on the foundational optical OFDM techniques to form a new optical OFDM signal. Asymmetrical clipped DC-biased optical OFDM (ADO-OFDM) described in \cite{da13} incorporates an ACO-OFDM component and a DCO-OFDM component. Hybrid asymmetrical clipped OFDM (HACO-OFDM) introduced in \cite{rk14} relies on a combination of an ACO-OFDM signal and a pulse amplitude modulated discrete multitone (PAM-DMT) signal. Asymmetrically and symmetrically clipped optical OFDM (ASCO-OFDM) introduced in \cite{wb15} relies on ACO-OFDM signals and a modified DCO-OFDM signal. Enhanced unipolar OFDM (eU-OFDM) introduced in \cite{th14} relies on multiple unipolar OFDM signals which in turn is based off DCO-OFDM.

In this paper, we describe, evaluate, and benchmark our technique: spectral and energy efficient OFDM (SEE-OFDM), initially introduced in \cite{el14a} and expanded by Wang et al. under a different name in \cite{Wang15}. SEE-OFDM builds upon ACO-OFDM and is a combination technique that combines different length ACO-OFDM signals together to create one SEE-OFDM signal. By combining multiple components together, SEE-OFDM can achieve a spectral efficiency similar to DCO-OFDM. Since SEE-OFDM builds off ACO-OFDM signals, it does not require a DC-bias, similar to ACO-OFDM. SEE-OFDM also has a lower peak-to-average-power rate (PAPR) than ACO-OFDM and the other combination methods. Low PAPR is desired in OWC IM/DD systems due to the linear range constraints on optical sources. New in this work are (1)comparisons between the separate SEE-OFDM components, (2) a method for generating SEE-OFDM completely in the frequency-domain, (3) comparisons among the different receiver methods: reconstruction, soft-decoding, hard-decoding, and (4) comparisons to HACO-OFDM, ASCO-OFDM, and eU-OFDM. 

This paper starts with a review in Section II of the IM/DD optical OFDM system as well as the current spectral and energy efficiency limitations regarding the foundational optical OFDM techniques: DCO-OFDM and ACO-OFDM. Section III follows with a  description and evaluation of SEE-OFDM. Other hybrid state-of-the-art techniques are discussed in Section IV, including HACO-OFDM, ASCO-OFDM, and eU-OFDM. Comparisons are made between our proposed SEE-OFDM against ACO-OFDM, as a baseline, and also against the other combination optical OFDM formats in Section V. Finally, conclusions are given in Section VI. 

\section{IM/DD Optical OFDM Systems}\label{OFDM in optical systems}
\begin{figure*}[t]
  \centering
  \includegraphics[scale=0.55]{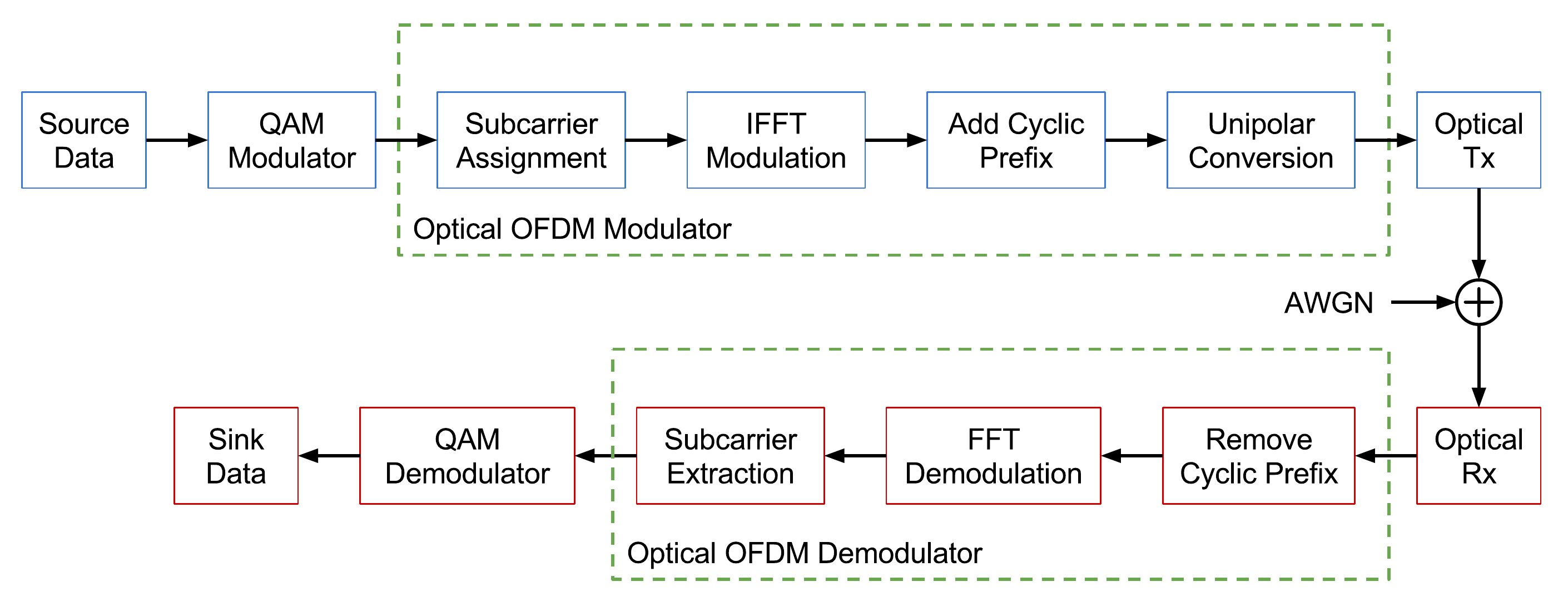}
  \caption{Block diagram of a typical IM/DD OFDM system.}\label{OWC_System}
\end{figure*}
Unlike RF-OFDM, IM/DD OFDM requires a real and positive baseband signal; there is no carrier frequency. This restriction to a real and positive signal fundamentally changes how we generate an optical OFDM signal. In Fig.~\ref{OWC_System}, a typical IM/DD OFDM system is shown. The OFDM box illustrates the steps required for optical OFDM. In OFDM, quadrature constellations are assigned to individual subcarriers in the frequency-domain. An inverse fast Fourier transform (IFFT) operation is then used to modulate these subcarriers, resulting in a time domain signal that can be transferred to an optical source. A cyclic prefix (CP) is usually added to prevent inter-carrier interference (ICI) prior to modulating the optical source \cite{we71}. Both ACO-OFDM and DCO-OFDM impose a Hermitian symmetry to ensure a real, time-domain signal. However, ACO-OFDM modulates only odd subcarriers \cite{al06} while DCO-OFDM modulates all subcarriers.

Next, the signals must be converted to unipolar signals. For DCO-OFDM, the entire time-domain signal is DC-biased. After DC-biasing, any remaining negative values are clipped to zero. The optimal DC-bias to apply is proportional to the standard deviation of the electrical signal \cite{bk11}. For ACO-OFDM, due to the antisymmetry, all data that exist on the negative values of the time domain signal also exist on the positive values of the signal. Therefore, clipping the redundant negative values to zero is possible with no noise on the data carrying odd subcarriers \cite{al06}.

By imposing Hermitian symmetry on the subcarrier mapping, the maximum bandwidth of an OFDM signal is constrained to half the bandwidth of RF-OFDM, as half the subcarriers are carrying redundant data, the conjugate of the first half. Therefore, half the bandwidth of RF-OFDM is the maximum spectral efficiency possible for Hermitian symmetry based optical OFDM. Because DCO-OFDM  modulates all available subcarriers, we will use it as a baseline modulation when comparing the spectral efficiency with other methods. On another note, polar-OFDM, an optical OFDM not based on Hermitian symmetry, has the same maximum spectral efficiency as DCO-OFDM \cite{el15}. Therefore, currently to the authors' knowledge, the maximum spectral efficiency of an IM/DD OFDM system is that of DCO-OFDM. The rate of DCO-OFDM is given below
\begin{equation} \label{DCO_rate}
R_{\textrm{DCO}} = \frac{N/2-1}{(N+N_{\textrm{CP}})}B\log_{2}M~bits/s
\end{equation}
where B is the bandwidth, $M$ is the quadrature amplitude modulation (QAM) modulation order, $N_\textrm{CP}$ is the number of samples used for the CP, and $N$ is the length of the IFFT. ACO-OFDM, on the other hand, only uses the odd subcarriers of the available subcarriers. This makes ACO-OFDM less spectrally efficient than DCO-OFDM by about half. The rate of ACO-OFDM is given 
\begin{equation} \label{ACO_rate}
R_{\textrm{ACO}} = \frac{N/4}{(N+N_{\textrm{CP}})}B\log_{2}M~bits/s
\end{equation}
For a fixed power allocation, for ACO-OFDM to have the same rate as DCO-OFDM, it requires a higher-order constellation and hence will have a higher bit error rate (BER) for the same signal-to-noise ratio (SNR). Therefore, for a fixed number of bits per a subcarrier, DCO-OFDM has a better spectral efficiency than ACO-OFDM.

However, the DC-bias in DCO-OFDM will also increase the energy consumption of the system. Although DCO-OFDM has a better spectral efficiency given a fixed number of bits per subcarrier, the energy drawbacks of DCO-OFDM make DCO-OFDM less attractive than ACO-OFDM.   DCO-OFDM requires an additional DC-bias on top of the DC-bias required for powering the optical source that significantly increases the required SNR for a given BER \cite{meh11b}. ACO-OFDM was created to combat the high energy requirements of DCO-OFDM \cite{al06}. As an energy-efficient technique, ACO-OFDM does not require an additional DC-bias. Instead, ACO-OFDM takes advantage of the antisymmetry in the time-domain signal created by only modulating the odd subcarriers \cite{al06}. This redundancy in the negative values make it possible to simply clip to zero the negative values. While, ACO-OFDM is more energy efficient than DCO-OFDM, it has less spectral efficiency than DCO-OFDM. Additionally, ACO-OFDM like other OFDM signals has a high PAPR. The DC-bias in DCO-OFDM increases the average power and so decreases the PAPR for DCO-OFDM. 

\section{SEE-OFDM}
SEE-OFDM combines multiple ACO-OFDM-like components together, achieving greater spectral and energy efficiency than ACO-OFDM or DCO-OFDM. Here, we describe the generation of SEE-OFDM in the frequency-domain and then show how that same signal can also be generated in the time-domain. Following, we discuss the spectral efficiency, data rate, and interference of SEE-OFDM. Discussion of the different methods for receiving SEE-OFDM are described afterward followed by a section evaluating the performance of SEE-OFDM.

\subsection{Transmitter}
\begin{figure*}[!htb]
  \centering
  \includegraphics[scale=0.66]{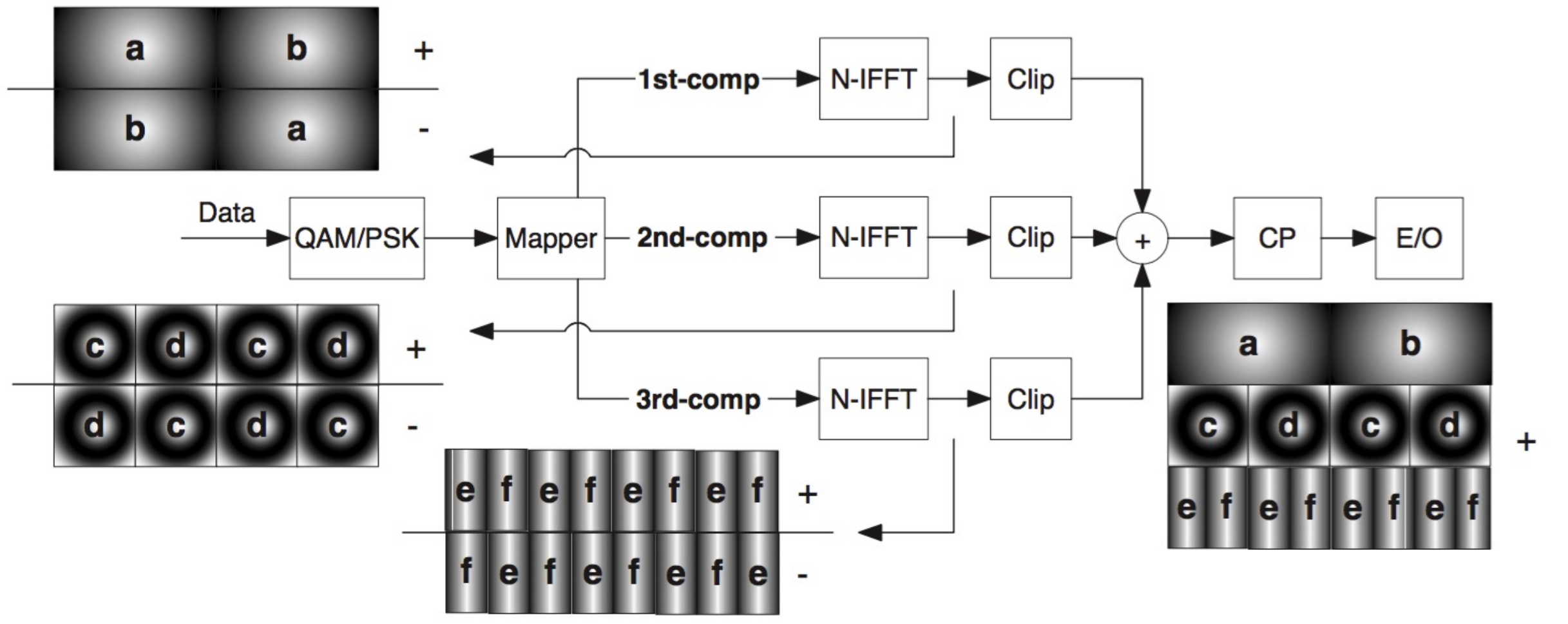}
  \caption{This figure demonstrates a three-component SEE-OFDM transmitter. The first-component is an ACO-OFDM signal using an $N$-point IFFT. The portion marked `a' represents the positive portion of the first half of the signal and `b' represents the positive portion of the second half of the signal. The first-component has anti-symmetry and therefore, `a' and `b' exist in the negative domain as well but on opposite halves of the signal. The subsequent components are $N/2$-length and $N/4$-length ACO-OFDM signals, with anti-symmetry and
then repetition.}\label{SEETX}
\end{figure*}
SEE-OFDM is a multi-component approach. The first component is a conventional ACO-OFDM signal. Given a maximum of $N$ subcarriers, the $n$-indexed time-domain samples of the first-component signal, $x_{\textrm{SEE},1}(n)$, are 
\begin{equation} \label{1st_component}
x_{\textrm{SEE},1}(n) = \left(\textrm{Real}(\sum_{k=0}^{N/4-1} X_{\textrm{SEE},1}(k)~e^{\frac{j2\pi (2k+1)n}{N}})\right)^+
\end{equation}
where the $k$-indexed $X_{\textrm{SEE},1}(k)$ are the frequency-domain input symbols (constellation values) associated with the first-component. The notation $\textrm{Real}(.)$ indicates a real output signal (samples) which can be implemented by imposing a Hermitian symmetry in the subcarrier assignment; while $(.)^{+}$ denotes clipping the signal at zero to realize a unipolar real and positive signal. The unclipped $x_{\textrm{SEE},1}(n)$ signal is asymmetric in the sense that $x_{\textrm{SEE},1}(n+N) = -x_{\textrm{SEE},1}(n)$. The second time-domain component $x_{SEE,2}(n)$ is constructed in the following way. Rather than modulate the odd subcarriers, we modulate the  even-odd subcarriers , $2(2k+1)$ for $k=0,\ldots, N/8-1$. It has the form:
\begin{equation} \label{2nd_component}
x_{\textrm{SEE},2}(n) = \left(\textrm{Real}(\sum_{k=0}^{N/8-1} X_{\textrm{SEE},2}(k)~e^{\frac{j2\pi 2(2k+1)n}{N}})\right)^+
\end{equation}
where the $k$-indexed $X_{\textrm{SEE},2}(k)$ are the frequency-domain input symbols (constellation values) associated with the second-component. Like the first component of SEE-OFDM, the unclipped $x_{\textrm{SEE},2}(n)$ signal is asymmetric in the sense that $x_{\textrm{SEE},2}(n+N/4) = -x_{\textrm{SEE},2}(n)$ and $x_{\textrm{SEE},2}(n+N/2) = x_{\textrm{SEE},2}(n)$.

A two-component signal $x_{\textit{SEE}}$ is equal to
\begin{equation} \label{2-comp}
x_{\textrm{SEE}} = x_{\textrm{SEE},1}+x_{\textrm{SEE},2}
\end{equation}
where the summation takes place after clipping the real and bipolar  $x_{\textrm{SEE},1}$ and $x_{\textrm{SEE},2}$ 
signals at zero to achieve unipolar positive signals.
\begin{figure}[!htb]
  \centering
  \includegraphics[scale=0.95]{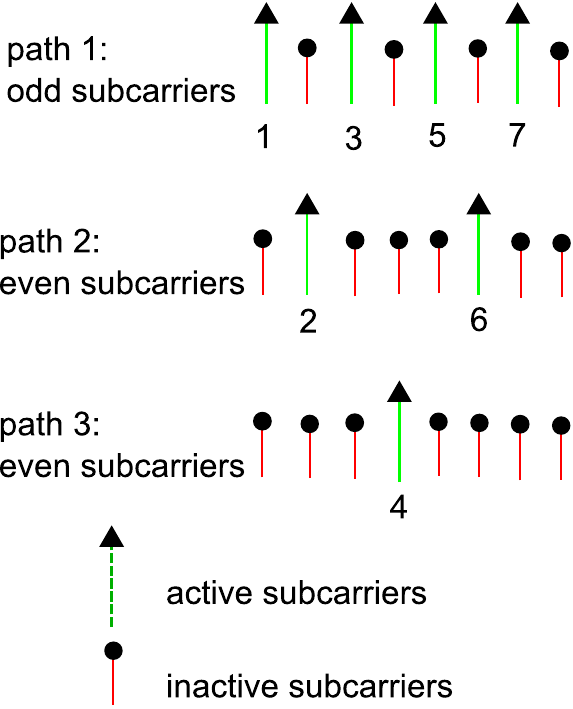}
  \caption{Active subcarriers of the three-component SEE-OFDM transmitter that carry data.}\label{SEETXFREQ}
\end{figure}
We can continue adding components where the number of active subcarriers per component, $N_{p}$, is 
\begin{equation} \label{np}
  N_{p}= \frac{N}{2^{p+1}}\\
\end{equation}
where $p=1,\cdots,\log_{2}(\frac{N}{2})$. For example, and considering $N=16$, $N_{1}=4$, $N_{2}=2$ and $N_{3}=1$ are the number of active subcarriers for the individual components, the indices of the active subcarriers per component $N_{p}^{s}$ are described as,
\begin{equation} \label{nps}
  N_{p}^{s}=2^{p-1}(2k+1)\\
\end{equation}
where $k=0,\cdots,[\frac{N}{2^{p+1}}-1]$. For example, and again considering $N=16$ and a three-components transmitter, $N_{1}^{s}=\{1, 3, 5, 7\}$, $N_{2}^{s}=\{2,6\}$ and $N_{3}^{s}=\{4\}$ are the indices of active subcarriers for the individual components (Fig.~\ref{SEETXFREQ}).

The $n$-indexed time sample of the $p^{th}$-component signal, $x_{\textrm{SEE},p}(n)$, is given by
\begin{equation} \label{pth-comp}
x_{\textrm{SEE},p}(n) = \left(\textrm{Real}(\sum_{k=0}^{{N/2^{p+1}}-1} X_{\textrm{SEE},m}(k)~e^{\frac{j2\pi 2^{p-1}(2k+1)n}{N}})\right)^+
\end{equation}
where the $k$-indexed $X_{\textrm{SEE},p}(k)$ are the frequency-domain input symbols associated with the $p^{th}$-component. 

An $r$-component signal $x_{\textit{SEE}}$ is equal to
\begin{equation} \label{r_comp}
x_{\textrm{SEE}} = x_{\textrm{SEE},1}+x_{\textrm{SEE},2}+ . . . +x_{\textrm{SEE},r}
\end{equation}
where $r$ is $\leq\log_{2}(\frac{N}{2})$. 
The additional time-domain components $x_{\textrm{SEE},p}$ do not interfere with the prior $p-1$ components. This is shown in Appendix~\ref{interference}. While each subsequent component does not interfere with the prior components, prior components do interfere with the subsequent components, which can be eliminated at the receiver.

Fig.~\ref{SEETX} shows the generation of a $3$-component SEE-OFDM in a flow diagram.  As seen in Fig.~\ref{SEETX}, the subsequent components of SEE-OFDM have anti-symmetry and repetition in the time-domain. For example, the second-component signal is actually an $N/2$-length ACO-OFDM signal with anti-symmetry repeated and the third-component signal is actually an $N/4$-length ACO-OFDM signal with anti-symmetry repeated three times. A proof of this is shown in Appendix~\ref{timefreq}. As a result, it is possible to generate the components of SEE-OFDM in the time-domain without selecting active subcarriers in the frequency-domain based on Eq.~\ref{np} but by using different length IFFTs. This was the original approach introduced in \cite{el14a}. 

\begin{figure*}[!htb]
  \centering
  \includegraphics[scale=0.66]{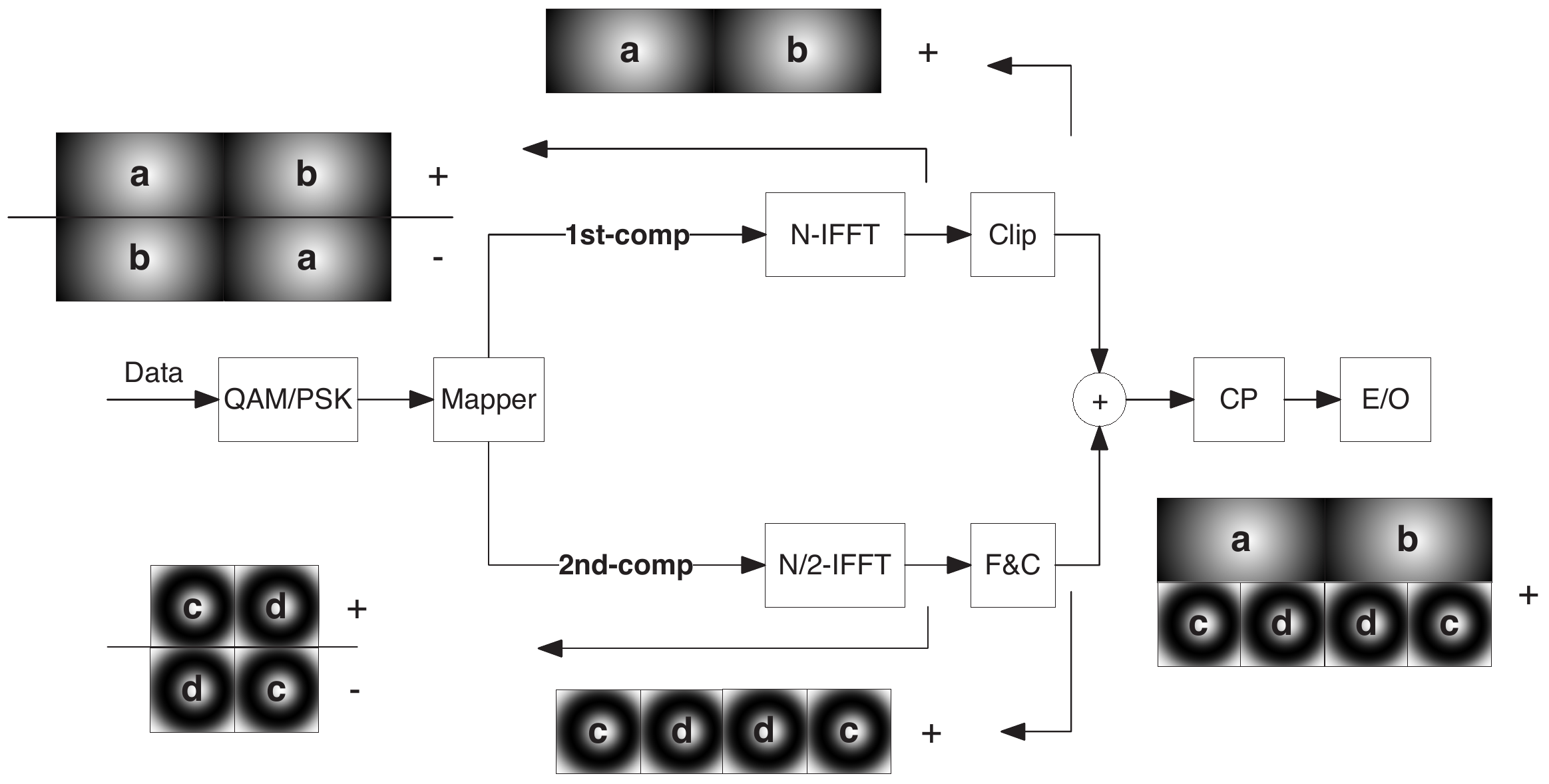}
  \caption{The two-component SEE-OFDM transmitter with the second component using an $N/2$ length IFFT. This has the same anti-symmetry and repetition in the time-domain signal as seen in the frequency-domain only generation using $N$-length IFFT for all components.}\label{SEETXtime}
\end{figure*}

Using an $N$-length IFFT, a conventional ACO-OFDM signal is generated as the first-component as seen in Fig.~\ref{SEETXtime}. Since this is a traditional ACO-OFDM signal, there is repetition in the negative portion of the signal due to the antisymmetry of the signal. As a consequence, the negative part can be clipped without any loss of information. Using different IFFT lengths, additional ACO-OFDM signals are generated for the subsequent components. The IFFT length is halved with each additional component to obtain the symmetry desired. As shown in Fig.~\ref{SEETX}, the same procedure is repeated in the second component using an $N/2$-length IFFT operation. After clipping, each additional signal is repeated $L$ times to match the length of the first-component, where $L=2^{(p-1)}$ and $p$ is the index of the component as defined above. As shown in Fig.~\ref{SEETXtime}, the different real-valued unipolar signals from the two different components are then added together.\footnote{SEE-OFDM can be generated in time or frequency-domain but one must choose scaling factors to ensure parity.} The relative scaling of the additional components is an important design configuration and will be explored in Section~\ref{SEE_results}.

\subsection{Spectral Efficiency, Data Rate, and Interference}\label{SEE_spec}

SEE-OFDM has a variable spectral efficiency. Depending on the number of components used to create the final SEE-OFDM signal, as seen in Eq.~\ref{r_comp}, the spectral efficiency changes. At a minimum the spectral efficiency is three-fourths of DCO-OFDM and at a maximum the spectral efficiency is equivalent to DCO-OFDM. The spectral efficiency $\eta_{\textrm{SEE}}$ in percentage of a SEE-OFDM transmission is 
\begin{equation} \label{transmission_spec}
  \eta_{\textrm{SEE}}=\frac{\sum_{p=1}^{p=r}N_{p}}{(N/2)}\times100\\
\end{equation}
For example, with $N=16$ subcarriers and a three-component signal, $\eta_{\textrm{SEE}}=87.5\%$. Ignoring the transmitter power, the
spectral efficiency for this specific case is equal to the spectral efficiency of a DCO-OFDM system, where the value of the first and the
$N/2$ subcarriers must be zero to ensure the Hermitian property.

The achieved data rate of SEE-OFDM is given by
\begin{equation} \label{SEE_rate}
\begin{split}
R_{\textrm{SEE}} &= R_{\textrm{SEE},1}+R_{\textrm{SEE},2}+ . . . +R_{\textrm{SEE},r}\\
&= \frac{\sum_{p=1}^{p=r}N_{p}}{(N+N_{\textrm{CP}})}B\log_{2}M~bits/s
\end{split}
\end{equation}
where $B$ is the bandwidth, $N_\textrm{CP}$ is the number of samples used for the CP, and $M$ is the QAM modulation order. $N_{p}$ is defined in Eq.~\ref{np}.

The periods of the time-domain signals are $N$, $N/2$ and $N/4$ for
the first, second and third components, respectively. Consequently,
the clipping operation on the signals of the second and third
components will not distort the constellation values on the active
subcarriers of the previous components. So the second component signal will not interfere
with the modulated subcarriers of the first component; the third component signal will not interfere with the modulated
subcarriers of the first and second components; and so on.
A proof of this is given in Appendix~\ref{interference}.  

\subsection{Receiver Methods}

There are two ways to decode SEE-OFDM: soft-decision decoding and hard-decision decoding. Soft-decision decoding was introduced in \cite{el14a} as a reconstruction technique. It is also shown in \cite{Wang15} using an iterative subtraction receiver. But as shown below, the hard-decision, iterative subtraction decoding is 1.5 dB better than soft-decision decoding (reconstruction or iterative subtraction) with very little extra complexity. In the following discussion of receiver methods, we assume that the receiver clips any negative noise as in \cite{wa09}. In an additive white Gaussian noise (AWGN) channel and for ACO-OFDM, clipping the receiver noise improves the performance through 1.25 dB increase in the effective SNR \cite{wa09}. We assume that clipping at the receiver for SEE-OFDM has a similar effect.

\subsubsection{Reconstruction}
\begin{figure}[!htb]
  \centering
  \includegraphics[scale=0.65]{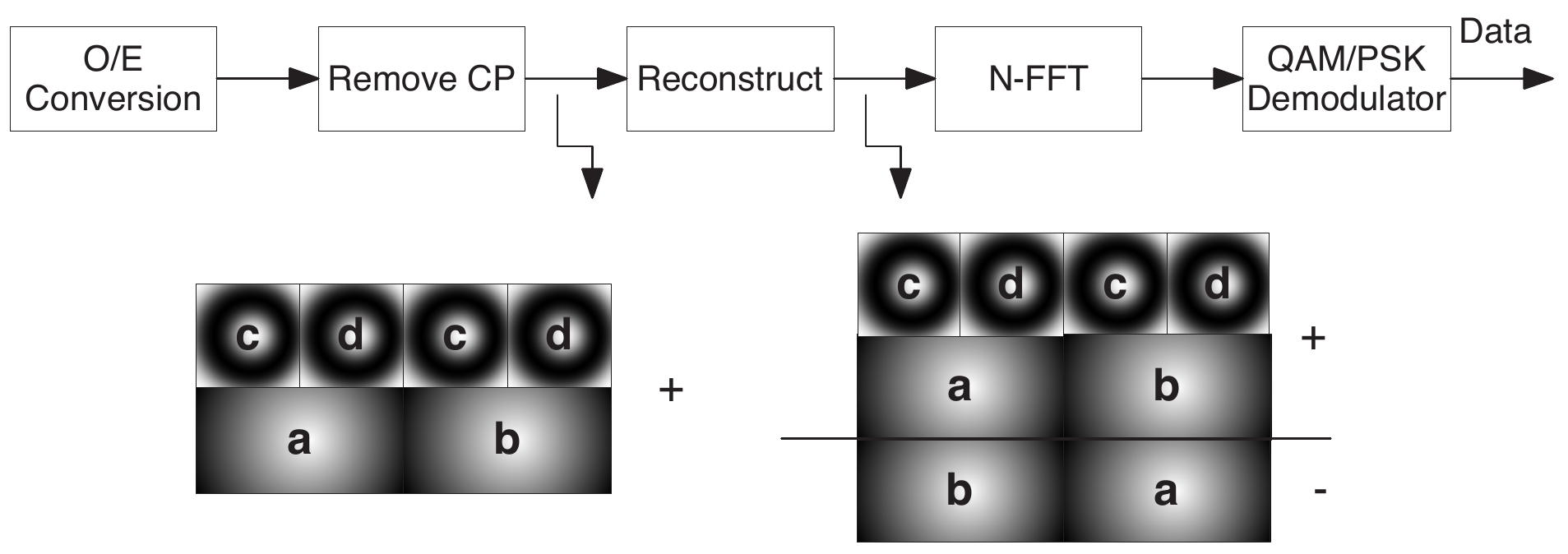}
  \caption{Two-components SEE-OFDM receiver showing the pre-reconstruction symmetry and the after construction symmetry that re-establishes the negative portions of the first component to eliminate the noise on the even subcarriers so a single FFT can be performed.}\label{SEERX}
\end{figure}

\begin{figure}[!htb]
  \centering
  \includegraphics[scale=0.5]{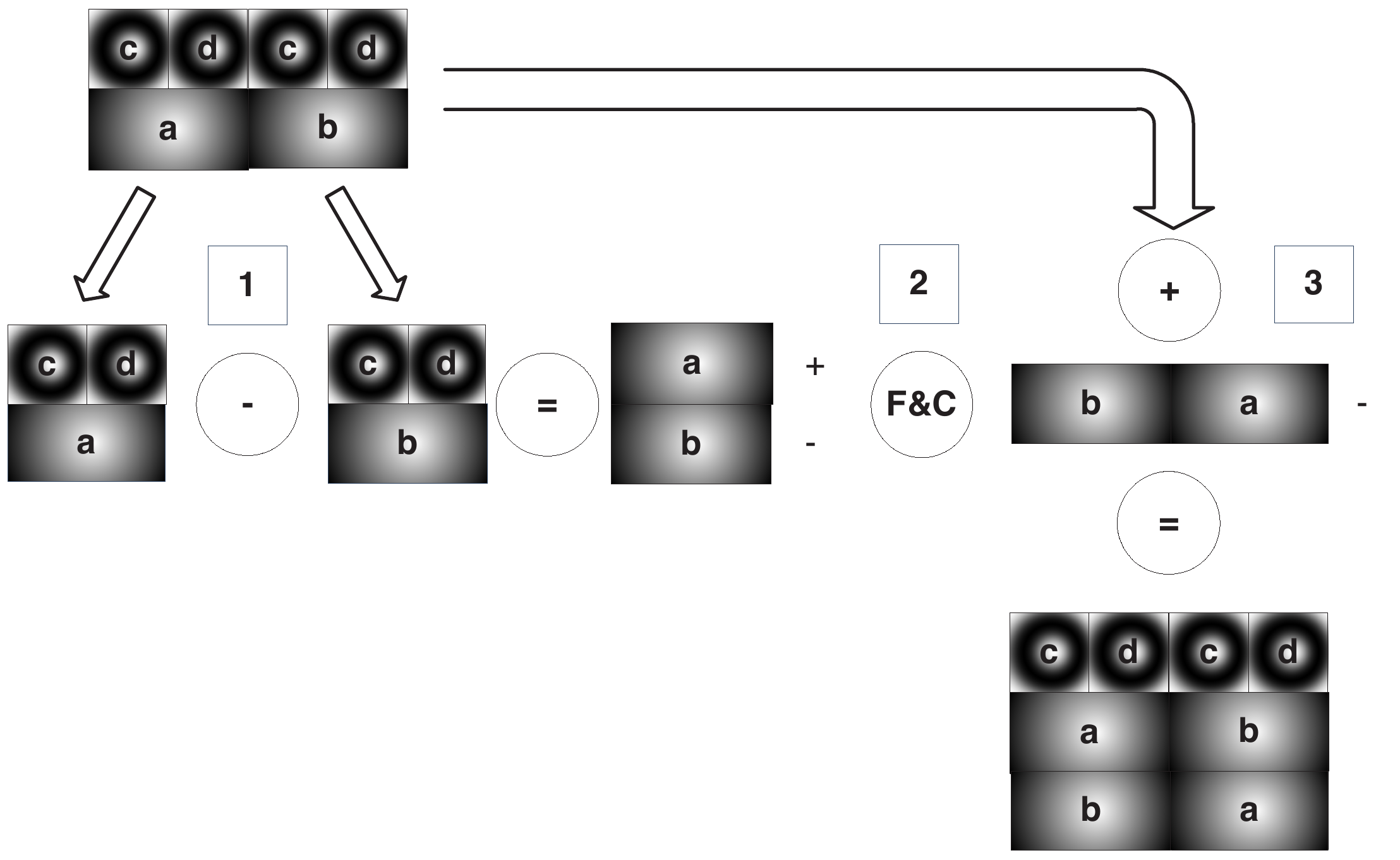}
  \caption{Two-components SEE-OFDM reconstruction steps to obtain the negative portion of the first component that was clipped at the transmitter.}\label{SEERXCON}
\end{figure}

\begin{figure}[!htb]
  \centering
  \includegraphics[scale=0.65]{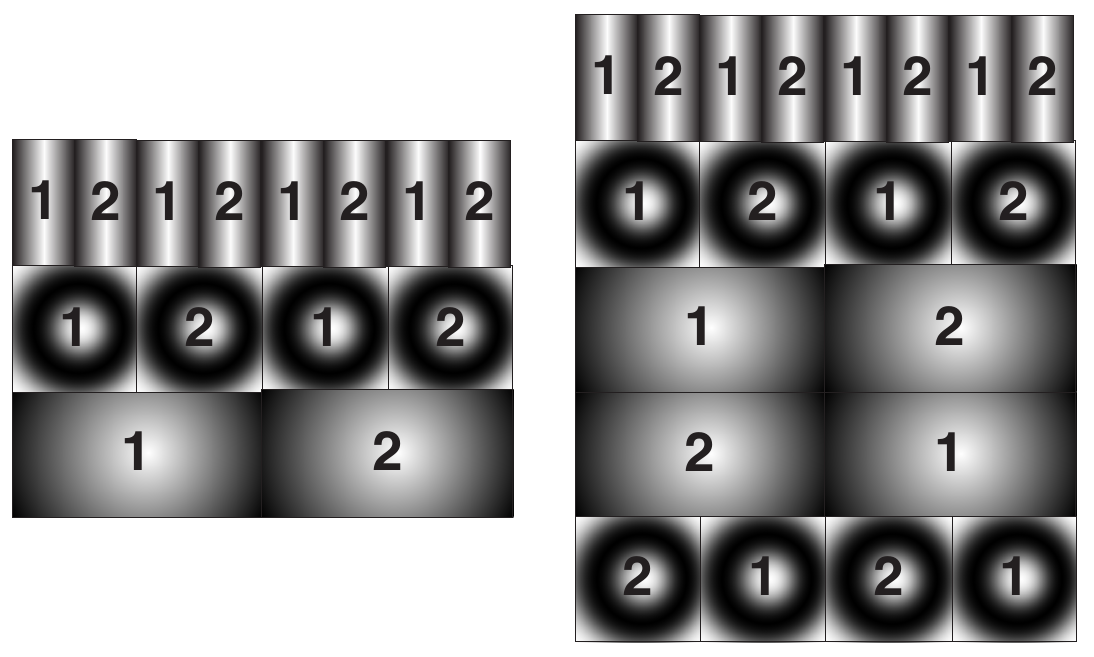}
  \caption{Three-components SEE-OFDM signal received at the receiver and the signal after it has been reconstructed to remove noise from prior components on the subsequent components.}\label{SEERXCON3}
\end{figure}

Reconstruction at the receiver for SEE-OFDM was introduced in \cite{el14a}. The building blocks of the SEE-OFDM receiver using the reconstruction method is shown in Fig.~\ref{SEERX}. A pre-conditioning step is required before applying a single $N$-length fast Fourier transform (FFT) operation. In such step, the original bipolar signal of the first component is reconstructed. The main purpose of this step is to eliminate the intermodulation caused by the signal clipping in the first component which occurs in the subsequent components due to clipping interference.
At the receiver, and after the optical-to-electrical conversion using an optical detector, \textit{i.e.} a photodiode (PD) and assuming an AWGN channel model, the time-domain samples can be expressed as follows:
\begin{equation}\label{channela}
y[n] = x[n] \otimes h[n] + z[n]
\end{equation}
where, $h[n]$ is the impulse response of the channel ($h[h]=\delta[n]$), $z[n]$ is the AWGN with variance $\sigma_{z}^{2}$, \textit{i.e.} noise power and $\otimes$ denotes a convolution operation. As shown in Fig.~\ref{SEERXCON}, a series of operations are required to reconstruct the negative portion of the first component. By reconstructing the negative portion of the signal, a single FFT operation using the bipolar signal of the first path will eliminate the clipping interference of the first component on the subsequent component. The first step in the reconstruction process is the subtraction of the second half-period $y_{2}[n]$ of $y[n]$ from the first half-period  $y_{1}[n]$ to obtain  $r_{a}[n]$.
\begin{eqnarray}\label{channelb}
 \nonumber
  r_{a}[n] &=& (y_{1}[n]-y_{2}[n]) + z_{1}[n] + z_{2}[n] \\
   &=& (y_{1}[n]-y_{2}[n]) + z[n]
\end{eqnarray}
where, $z_{1}[n]$ is the AWGN during $y_{1}[n]$ and $z_{2}[n]$ is the AWGN during $y_{2}[n]$ and $z[n]=z_{1}[n]+z_{2}[n]$ denotes the sum Gaussian noise which has the power $\sigma_{z}^{2}$.
The second step is flipping the polarity of the negative samples of $r_{a}[n]$ followed by a horizontal concatenation with the positive samples of $r_{a}[n]$ to form a length $N$ time-domain symbol $r_{b}[n]$.
\begin{equation}\label{channelc}
r_{b}[n] = (-r_{a}^{+}[n] \parallel r_{a}^{-}[n]) + z[n]
\end{equation}
where $r_{a}^{+}$ represents the positive samples of $r_{a}[n]$, $r_{a}^{-}$ represents the negative samples of $r_{a}[n]$ and $(\parallel)$ denotes the concatenation operator.
The last step to obtain the input signal to the FFT operation $r[n]$ is the summation of $y[n]$ and $r_{b}[n]$.
\begin{equation}\label{channeld}
r[n] = (y[n]+ r_{b}[n]) + z[n] + z[n]
\end{equation}
where, $z[n]+z[n]$ denotes the sum Gaussian noise which has the power $2\sigma_{z}^{2}$. The noise power of the SEE-OFDM is doubled during the reconstruction of the signal before the FFT operation. In a single-component (conventional ACO-OFDM), since there is no reconstruction, the noise power is $\sigma_{z}^{2}$, \textit{i.e.} half of the amount in SEE-OFDM. The equivalent SNR per received sample for two-components SEE-OFDM is given by,
\begin{equation}\label{channele}
\textrm{SNR} = 10\log_{10}\frac{{\cal E}_x}{2\sigma_{z}^{2}} 
\end{equation}
where, ${\cal E}_x$ denotes the transmitted signal power.
The same reconstruction procedure explained above for a two-component SEE-OFDM system is valid for a three-component SEE-OFDM system and so on. Fig.~\ref{SEERXCON3} shows the signal modulating the LED and the reconstructed signal at the receiver before applying the FFT for a three-component implementation.

The receiver decodes the time-domain OFDM symbol $x_{n}$ by performing the FFT operation,
\begin{equation} \label{ffta}
  X_{k}= \sum_{n=1}^{N} x_{n}\exp^{\left({-j\frac{2\pi}{N}nk}\right)}
\end{equation}
As mentioned in Appendix~\ref{interference}, there is no interference from prior components on the subsequent components. Therefore, when the FFT is performed over two consecutive and identical OFDM symbols (see the signal symmetry from the second component in Fig.~\ref{SEETX} and Fig.~\ref{SEERX}), the output of the FFT can be written as,
\begin{equation} \label{fftb}
  X_{k}= \sum_{n=1}^{N/2} x_{n}\exp^{\left({-j\frac{2\pi}{2N}nk}\right)}+\sum_{n=N/2+1}^{N} x_{n}\exp^{\left({-j\frac{2\pi}{2N}nk}\right)}
\end{equation}  

\subsubsection{Iterative Subtraction}
Another method of retrieving the data from the different components is through the use of iterative subtraction, which is a common receiver method among combination optical OFDM techniques \cite{da13,rk14,wb15,th14,Wang15}. In this section, we apply it to SEE-OFDM and discuss two ways to implement it: soft-decision and hard-decision decoding. The hard-decision, iterative receiver is different from soft-decision, iterative receiver in that it uses decided constellation values as opposed to estimated constellation values during the iterative decoding. To the authors' knowledge, hard-decision, iterative subtraction has not been studied for use with combination optical OFDM formats. Instead, soft-decision decoding is the method used in previous literature for other combination OFDM techniques \cite{da13,rk14,wb15,th14,Wang15}.

In the previous section and in \cite{el14a}, the received signal
goes through a reconstruction and conditioning stage before taking a
single FFT to recover the constellation values. This receiver method
increases the effective noise power by a factor that depends on the
number of components in the signal\cite{el14a}. However, because
$x_{\textrm{SEE},p}$ signals do not interfere with the prior
$x_{\textrm{SEE},1},\cdots,x_{\textrm{SEE},(p-1)}$ signals, successive
demodulation and interference cancellation can successfully mitigate the interference on the latter
components.  Given the received value $
y_{\textrm{SEE}}(n) = x_{\textrm{SEE}}(n)+w(n)$ where $w(n)$ is an
AWGN process, the successive demodulation and interference cancellation sequence is described below in
Algorithm 1.

\begin{algorithm}
\caption{Successive demodulation algorithm}
\label{DeModalgorithm}
\begin{algorithmic}[1]
\State $p=1$
\Repeat
\State FFT the time-domain signal $y_{\textrm{SEE},p}$.
\State Remove the effect of the channel using frequency-\par 
			 equalization.
\State Demodulate the constellation values of the $N_{p}^{s}$\par
			 subcarriers.
\State Let $y_{itt,p}(n),n=0,\cdots,N-1$ be the IFFT of the\par
			 equalized constellation values $N_{p}^{s}$.
\State Subtract $y_{itt,p}(n)$ from $y_{\textrm{SEE},p}$.
\State Set $p=p+1$ 
\State Save the output of the \textit{Step-7} subtraction in $y_{\textrm{SEE},p}$.
\Until all active subcarriers on individual components are demodulated.
\end{algorithmic}
\end{algorithm}

In summary, the basic idea is to reconstruct  the prior signal, $x_{\textrm{SEE},1},x_{\textrm{SEE},2},\ldots$ of the SEE-OFDM signal and subtract it from the received signal $y_{\textrm{SEE}}(n)$ to cancel the interference.  One can use either ``hard'' or ``soft'' subtraction. Hard-decision implies that a firm decision is made on the constellation values before reconstruction while soft-decision subtraction implies that the noisy constellation values are used to reconstruct the constituent signal. 

Given the demodulated data $\hat{Y}_{\textrm{SEE},p}(k)$ on the subcarriers with indices $N_{p}^{s}$, where the $\hat{Y}_{\textrm{SEE},p}(k)$ notation indicates that the receiver has made a decision on the received constellation value, this data can be remodulated and subtracted.

First assuming no errors in the demodulation of $\hat{Y}_{\textrm{SEE},p}(k)$, for $k\in N_{p}^{s}$,
\begin{flalign} \label{iterative_sub}
\quad y_{\textrm{SEE},p+1}(n) &= y_{\textrm{SEE},p}(n)-y_{itt,p}(n)+w(n) &
\end{flalign}
\begin{multline} \label{hard_sub}
y_{\textrm{SEE},p+1}(n) = y_{\textrm{SEE},p}(n)\\
-\left(\textrm{Real}(\sum_{k=0}^{N/(4\times 2^{p-1})}{\hat{Y}_{\textrm{SEE},p}(k)}~e^{\frac{j2\pi 2^{p-1}(2k+1)n}{N}})\right)^+ \\
+w(n)
\end{multline}

In other words, if there are no errors in demodulating the data, each successive contributing component can be demodulated with the same SNR as the prior constituent component. An inherent problem with hard-decision subtraction is that there can be error propagation. However, this can be mitigated by using error control codes on each constituent signal component, though as the number of subcarriers decreases, using an error control code on a low number of constellation values is impractical. However, we can use error-control codes that span several SEE-OFDM signals and decode a frame of several OFDM signals in parallel.

As will be seen later in the simulation section, hard-decoding without error-codes still performs better than the reconstruction and soft-decision subtraction receiver. With hard-decoding, if one or two errors in demodulation occur, the errors are spread over several time-domain samples. But more importantly, the goal of subtracting the reconstructed prior component from the current, is to subtract the the spillage from the clipping in the time-domain.  This spillage has a much smaller value than the constellation values, so one or two incorrect decisions may not corrupt the signal significantly.

\subsection{SEE-OFDM Performance Evaluation} \label{SEE_results}
For the simulation results presented below, an $N=64$-length IFFT/FFT  signal is used. The average electrical SEE-OFDM signal power ranges from  -4 dBm to 30 dBm. At the receiver, shot and thermal noise are modeled as AWGN  with an average power of -15 dBm (typical optical receiver sensitivity). Accordingly, the system SNR is in the range of 11 dB to 45 dB, which is in the acceptable reported range for indoor VLC systems. Moreover, before applying signal power, all values greater than a value of 1 of the signal, which is normalized to an average power of 1 W, is clipped to 1. This is to show the influence of the limited dynamic-range of the optical source and the associated induced clipping noise power is included in the model \cite{meh11b}. Perfect synchronization between the transmitter and the receiver is assumed. To insure a fair comparison, (1) the un-coded QAM constellations are assigned  so that all the compared methods have the same data rate  and (2) the average power calculated over one time-domain SEE-OFDM symbol of length $N$ are equal. A line-of-sight (LOS) configuration is assumed, thus no samples are used for the CP, \textit{i.e.} $N_{\textrm{CP}}=0$.

\begin{figure}[!htb]
  \centering
  \includegraphics[scale=0.65]{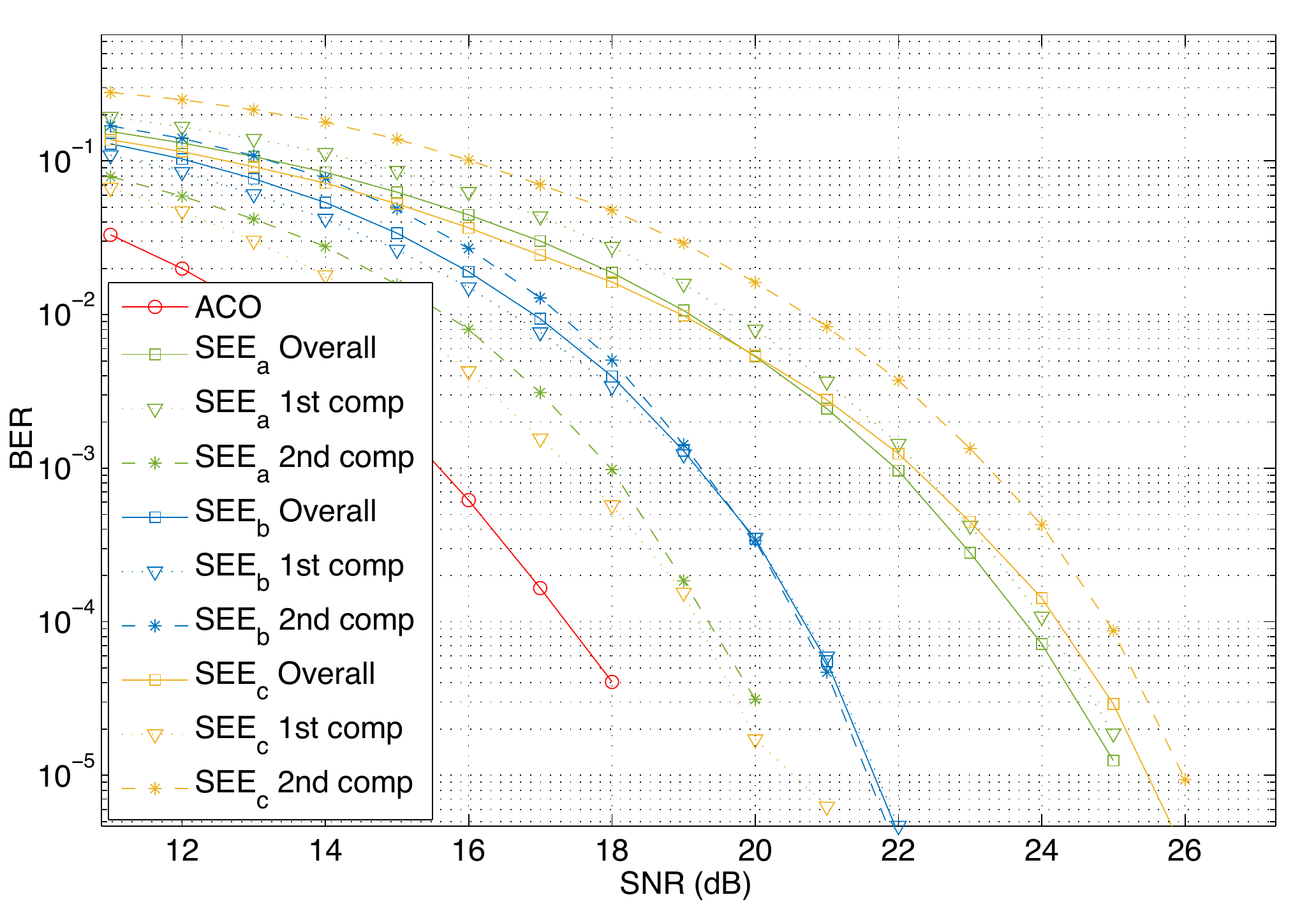}
  \caption{BER performance of the individual components of SEE-OFDM and overall SEE-OFDM signal bench-marked against ACO-OFDM at $16$-QAM with a hard-decision subtraction. Since all signals here have the same modulation order, the SEE-OFDM overall signals have higher bit rates than the ACO-OFDM signal. $SEE_{a}$ has more energy on the second-component, $SEE_{b}$ has equal energy allocation, and $SEE_{c}$ has more energy on the first-component. $SEE_{b}$, equal energy allocation, has the best overall performance of the three power allocation methods studied.}\label{diffpaths}
\end{figure}
First, the energy allocation of the SEE-OFDM transmitter is examined. Important to note is that using more than one component does indeed make the individual first component, the conventional ACO-OFDM signal component, of SEE-OFDM perform worse than just an ACO-OFDM signal. This is because for a single component ACO-OFDM signal, all signal power is allocated to that one component, whereas power is split amongst the different components for SEE-OFDM. It is expected that the first component of SEE-OFDM does not perform as well as ACO-OFDM signal. For the same $M$-QAM modulation order, SEE-OFDM has a higher BER than ACO-OFDM; but ACO-OFDM has an overall lower data rate due to the lower number of subcarriers used. This is explored further in Section \ref{comp_all}. In Fig.~\ref{diffpaths}, the performance of the different components of a 2-component SEE-OFDM implementation are simulated. SEE-OFDM with equal power allocation is called $\textrm{SEE}_{\textrm{b}}$. With $\textrm{SEE}_{\textrm{a}}$, the amplitudes of the $\textrm{2}^{nd}$-component are scaled by a factor of two of the baseline, meaning when normalized to an average signal power, the second-component will have more power. And finally, $\textrm{SEE}_{\textrm{c}}$ refers to a time-domain signal, where the amplitudes in the second-component are scaled by a factor of 1/2 of the baseline, resulting in less power in the second component when normalized to an average signal power. The QAM modulation order is kept constant at 16-QAM, the noise is kept constant at -15 dBm whereas the hard-decision subtraction receiver is used throughout. The single component ACO-OFDM in Fig.~\ref{diffpaths} is plotted as a baseline gauge. 

When more power is allocated to the $\textrm{2}^{nd}$-component, as in $\textrm{SEE}_{\textrm{a}}$, the second component performs, as expected, significantly better than the first component. The overall performance of both components performs in the middle of the two components and trends closer to the poorer first component performance. On the other hand, when less power is allocated to the second component, as in $\textrm{SEE}_{\textrm{c}}$, the second component performs poorly and the first component performs better. Again, the overall result trends closer to the poor performing component. For the $\textrm{SEE}_{\textrm{b}}$, the allocation of power between the two components is fairly equal, and for both components, the performance is similar. The overall performance of $\textrm{SEE}_{\textrm{b}}$ at a BER of $10^{-4}$ performs 3 dB better than the overall performances of the two unequal power allocations methods. 
\begin{figure}[!htb]
  \centering
  \includegraphics[scale=0.65]{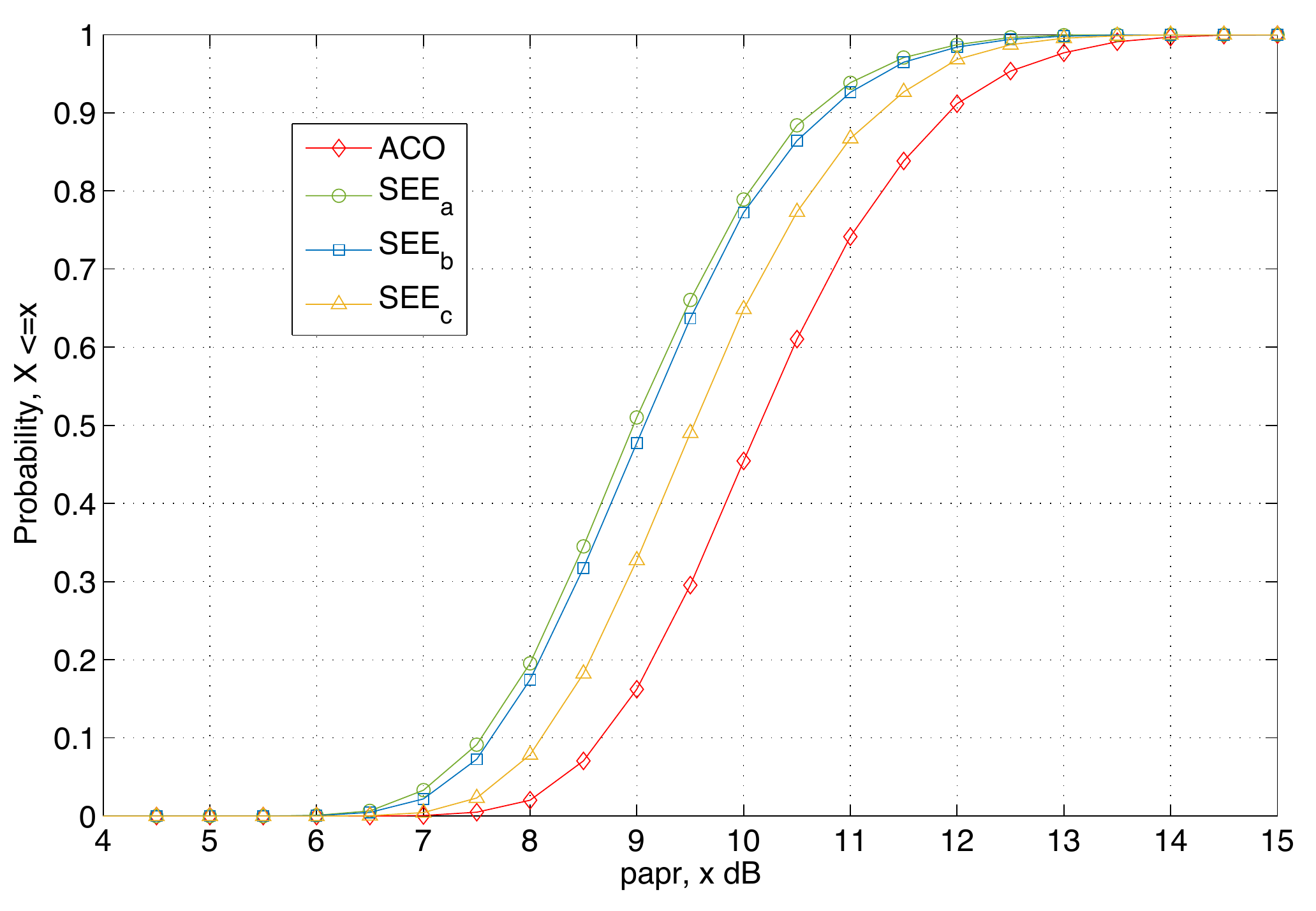}
  \caption{PAPR performance on individual components with different energy allocations. $SEE_{a}$ has more energy on the second-component, $SEE_{b}$ has equal energy allocation, and $SEE_{c}$ has more energy on the first-component. $SEE_{a}$ has the lowest PAPR.}\label{diffpaths_PAPR}
\end{figure}
However, when considering PAPR, more power on the second component is preferable. SEE-OFDM has a lower PAPR than ACO-OFDM because when different component signals add, there is more signal power which increase the average power, but not necessarily the peaks. When there is less power on the second component, the signal bears more resemblance to ACO-OFDM as shown in Fig.~\ref{diffpaths_PAPR}, where the $\textrm{SEE}_{\textrm{c}}$ curve is closer to the ACO-OFDM curve. With more power on the second component, $\textrm{SEE}_{\textrm{a}}$, the SEE-OFDM signal has the lowest PAPR. And with equal power on both components, $\textrm{SEE}_{\textrm{b}}$, the PAPR is in between the two. For a two-component SEE-OFDM implementation, the gap between the $\textrm{SEE}_{\textrm{a}}$ and $\textrm{SEE}_{\textrm{b}}$ is much more obvious, with $\textrm{SEE}_{\textrm{a}}$ having a much lower PAPR. However, PAPR does not tell a full story and the BER performance is a better performance indicator. A low PAPR can have bad BER performance.
\begin{figure}[!htb]
  \centering
  \includegraphics[scale=0.65]{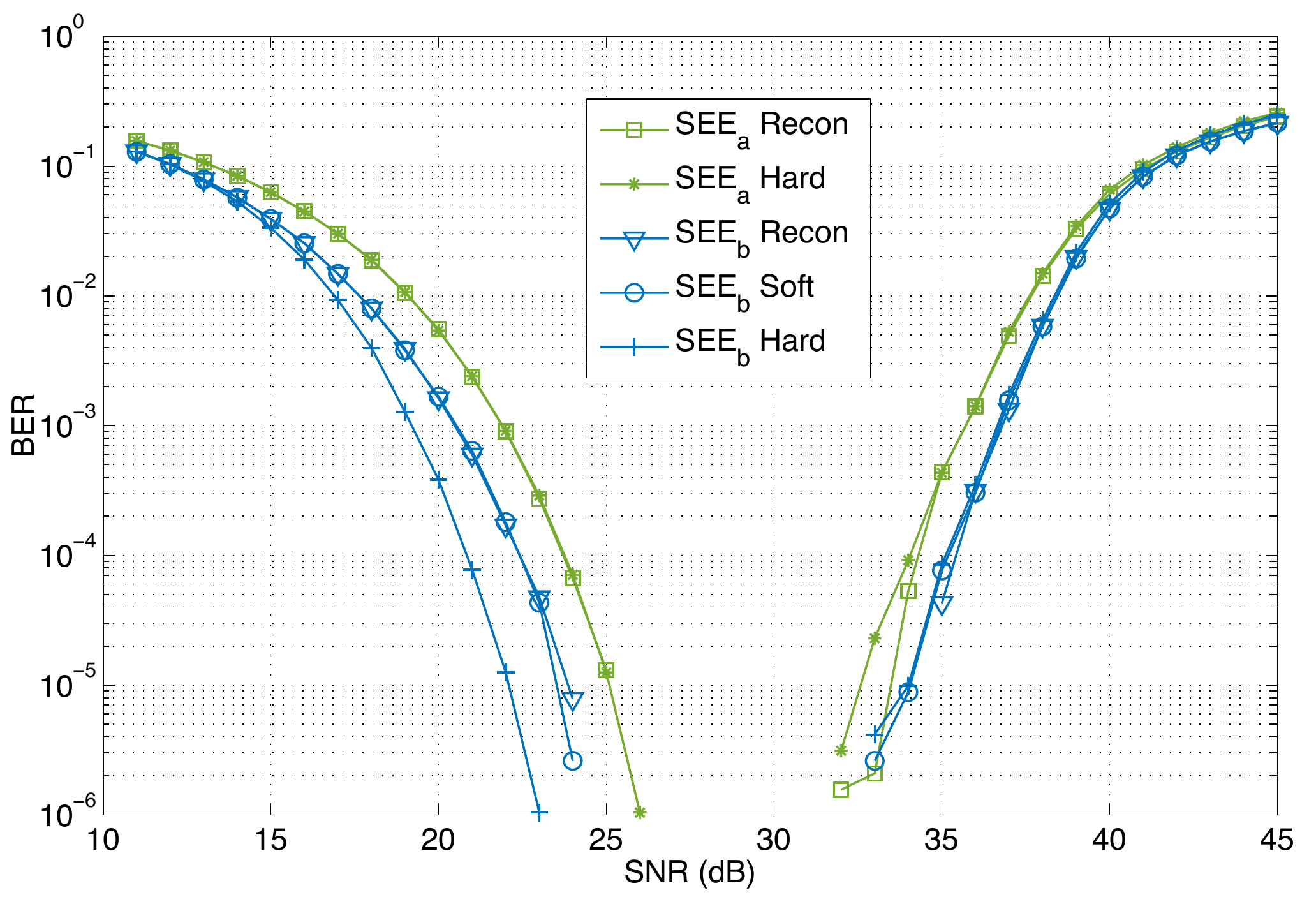}
  \caption{BER for the two-component SEE-OFDM transmitter comparing different energy allocations and receiver methods at $16$-QAM. The increase in BER is due to constricting the signal to a linear region. The left-most curve, $SEE_{b}$, with Hard Decoding has the lowest BER for the same rate.}\label{difftxrx}
\end{figure}

Next, the different receivers are investigated. The performance of different receivers methods (reconstruction, hard-decision subtraction, and soft-decision subtraction) is shown in Fig.~\ref{difftxrx}. Interestingly, the BER performance of $\textrm{SEE}_{\textrm{a}}$ is comparable regardless of the receiver method. However, with $\textrm{SEE}_{\textrm{b}}$, the ideal power allocation, and targeting a $10^{-4}$ BER, significant differences can be seen. The hard-decision subtraction receiver performs about 1.5 dB better than the soft-decision and reconstruction methods. Successive demodulation only outperforms the reconstruction method when using a hard-decision subtraction receiver as soft-decision subtraction performs similarly to reconstruction. By remodulating the decided constellation values, the hard-decision subtraction receiver does not suffer the noise enhancement issue that the soft-decision subtraction receiver does, hence the increase performance of the hard-decision subtraction receiver over the soft-decision subtraction receiver. As the SNR increases, the BER will eventually increase  due to clipping noise from the electronics. The best power allocation and receiver combination, $\textrm{SEE}_{\textrm{b}}$ and hard-decision subtraction, performs 3 dB better than the worst power allocation and receiver combination, $\textrm{SEE}_{\textrm{a}}$ and hard-decision subtraction. 

\section{Other Hybrid Techniques}
\begin{figure*}[!htb]
  \centering
  \includegraphics[scale=0.50]{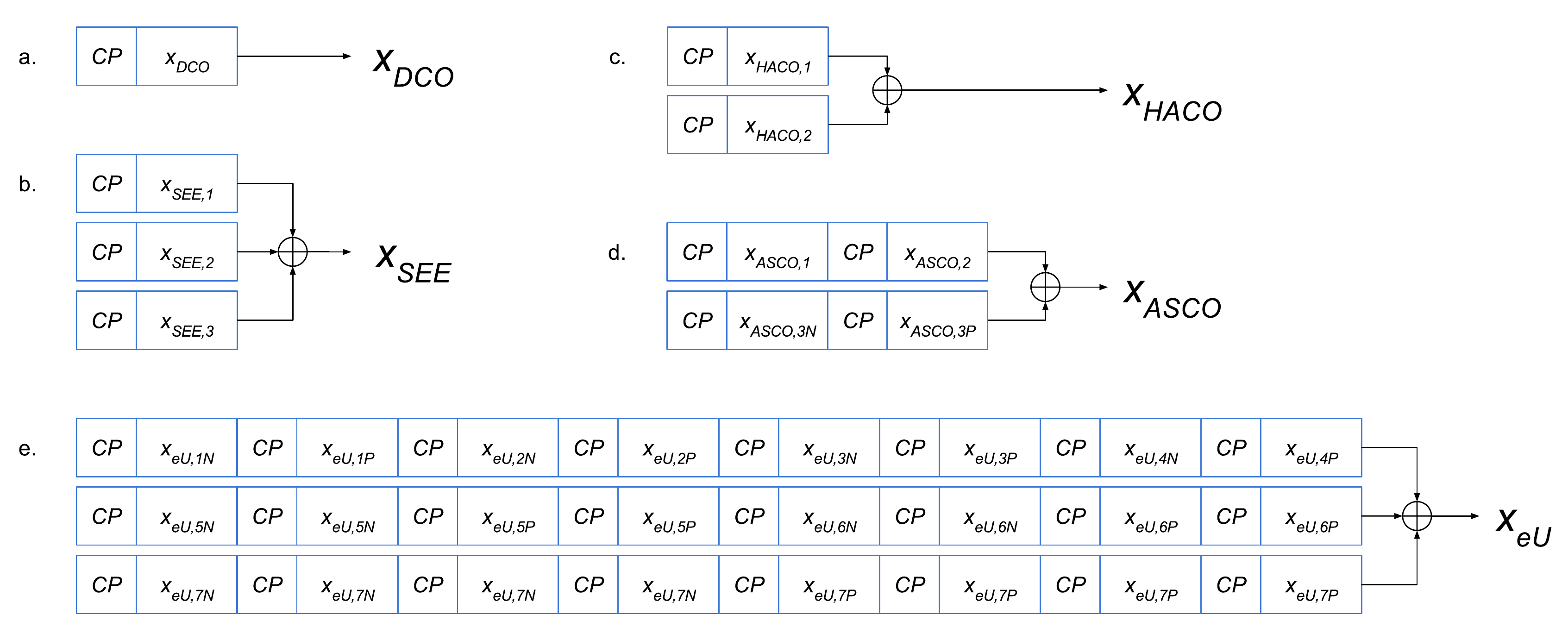}
  \caption{Frame construction of optical OFDM demonstrating the frame complexity required of each technique, where a frame is one OFDM symbol with length N. a. DCO-OFDM requires one frame, b. SEE-OFDM requires one frame regardless of number of components, c. HACO-OFDM requires two frames, d. ASCO-OFDM requires two frames, and e. eU-OFDM requires $2^{r}$, where r is number of components. In this diagram, all signals are clipped according and unipolar.}\label{frames}
\end{figure*}
To determine how well SEE-OFDM performs, we compare it to existing hybrid IM/DD methods. For example, in SEE-OFDM, different components consisting of ACO-OFDM modulated signals are combined together. The techniques considered in this section are unipolar techniques that do not require a DC-bias, which is why asymmetrically clipped DC-biased optical OFDM (ADO-OFDM) is not considered \cite{da13}.

\subsection{HACO-OFDM}
Hybrid asymmetrically clipped optical OFDM (HACO-OFDM) is introduced in \cite{rk14}. Here, we will give a brief overview of the technique. HACO-OFDM differs from SEE-OFDM in that it only and always uses two components: one ACO-OFDM signal and one PAM-DMT signal. The ACO-OFDM component is generated like a traditional ACO-OFDM signal, where a 2D mapping scheme, such as $M$-QAM, is used. Subcarriers are assigned as follows:
\begin{equation} \label{HACO1}
  X_{\textrm{HACO,1}} = [0, X_{\textrm{0}}, ..., 0, X_{\textrm{N/4-1}}, 0, X_{\textrm{N/4-1}}^*, 0 , ..., X_{\textrm{0}}^*]
\end{equation}
where $X_k$ are the constellation QAM symbols placed on $N/4$ of the subcarriers and $N$ is the total number of the subcarriers. A Hermitian symmetry is imposed similar to ACO-OFDM and DCO-OFDM. The time-domain output signal of the first component, $x_{\textrm{HACO,1}}$, is obtained by taking an $N$-length IFFT of the input vector. This component is a conventional ACO-OFDM signal. The second component is mapped using a 1D scheme such as $M$-PAM that will modulate just the imaginary parts of the even subcarriers. The input to the IFFT block is as follows:
\begin{equation} \label{HACO2}
  X_{\textrm{HACO,2}} = [0, 0, X_{\textrm{0}}, ..., 0, X_{\textrm{N/4-2}}, 0, 0, 0, X_{\textrm{N/4-2}}^*, 0 , ..., X_{\textrm{0}}^*, 0]
\end{equation}
This is different from a conventional PAM-DMT since only the even subcarriers, compared to all the subcarriers, are modulated. The time-domain output signal of the second component, $x_{\textrm{HACO,2}}$, is obtained by taking an $N$-length IFFT of the $X_{\textrm{HACO,2}}$. The negative parts of both time-domain components are then zero clipped and added together to form the HACO-OFDM time-domain signal, 
\begin{equation} \label{HACO}
x_{\textrm{HACO}} = (x_{\textrm{HACO,1}})^+ + (x_{\textrm{HACO,2}})^+
\end{equation}
as seen in Fig.~\ref{frames}-c. The signal clipping on the ACO component creates interference on the even subcarriers and the clipping on the PAM-DMT component creates interference on the real parts of the even subcarriers \cite{al06,lrbk09}. In both cases of clipping, no clipping noise is added to the odd subcarriers. However, the noise on the even subcarriers will cause the PAM-DMT performance to deteriorate \cite{rk14}. 

At the receiver, the first component is recovered first by performing an FFT and then selecting the odd subcarriers. Due to the clipping, the average power of the first component is halved, therefore the signal is multiplied by a factor of 2. Recovering the first component allows for regenerating the first component and predicting the noise on the even subcarriers. This is achieved similarly to the iterative subtraction method in SEE-OFDM. Once the second component is retrieved, the second component can be extracted and the PAM symbols can be recovered from the imaginary parts of the even subcarriers.

The overall bit rate that is achieved by this method is equivalent to the rate of DCO-OFDM, given as follow:
\begin{equation} \label{HACO_rate}
R_{\textrm{HACO}} = \frac{N/2-1}{(N+N_{\textrm{CP}})}B\log_{2}M~bit/s
\end{equation}
However, it is important to note, that one component of the HACO-OFDM uses 1-D PAM modulation, which is less energy efficient than an equal order QAM modulation \cite{rk14}.

\subsection{ASCO-OFDM}
Asymmetrically and symmetrically clipping optical OFDM (ASCO-OFDM) introduced in \cite{wb15} requires three IFFT and two frames with separate CP for each frame, where a frame is one OFDM symbol with length $N$, note Fig.~\ref{frames}-d. The first two IFFT modulates a conventional ACO-OFDM mapping, with subcarriers vectors assignment
\begin{equation} \label{ASCO1}
  X_{\textrm{ASCO,1}} = [0, X_{\textrm{0}}, ..., 0, X_{\textrm{N/4-1}}, 0, X_{\textrm{N/4-1}}*, 0 , ..., X_{\textrm{0}}*]
\end{equation}	
\begin{equation} \label{ASCO2}
  X_{\textrm{ASCO,2}} = [0, X_{\textrm{0}}, ..., 0, X_{\textrm{N/4-1}}, 0, X_{\textrm{N/4-1}}*, 0 , ..., X_{\textrm{0}}*]	
\end{equation}
Taking the IFFT of each of these two vectors would respectively give you $x_{\textrm{ASCO,1}}$ and $x_{\textrm{ASCO,2}}$. To ensure unipolarity of the signal, the negative values, for these two time-domain signals, can be zero clipped, since they are conventional ACO-OFDM signals and there no information lost due to the anti-symmetry. The third IFFT modulates a vector with the even subcarriers mapped with data and constrained by Hermitian symmetry.
\begin{equation} \label{ASCO3}
  X_{\textrm{ASCO,3}} = [0, 0, X_{\textrm{0}}, ..., 0, X_{\textrm{N/4-2}}, 0, 0, 0, X_{\textrm{N/4-2}}^*, 0 , ..., X_{\textrm{0}}^*, 0]
\end{equation}

Unlike, the first two component, the result from this IFFT, $x_{\textrm{ASCO,3}}$, cannot be simply clipped without lost of information. Instead, two signals are created from $x_{\textrm{ASCO,3}}$. $x_{\textrm{ASCO,3N}}$ is the $x_{\textrm{ASCO,3}}$ signal with the negative values zero-clipped and $x_{\textrm{ASCO,3P}}$ is the $x_{\textrm{ASCO,3}}$ signal with positive values zero-clipped followed by an absolute value operation. The ASCO-OFDM signal is then formed as seen in Fig.~\ref{frames}-d, consisting of two consecutive frames: $x_{\textrm{ASCO}}^{Frame1} = (x_{\textrm{ASCO,1}})^+ + x_{\textrm{HACO,3N}}$  and $x_{\textrm{ASCO}}^{Frame2} = (x_{\textrm{ASCO,2}})^+ + x_{\textrm{HACO,3P}}$. Each frame requires its own CP.

On the receiving end, the two frames are transformed to the frequency-domain through the use of an FFT, respectively. The odd subcarriers are selected out for the first and second component and the symbols are recovered. Again, similar to SEE-OFDM, these symbols are remodulated using an IFFT operation taken for both components and then respectively subtracted from Frame 1 and Frame 2 leaving just the third component data behind. By subtracting Frame 2, the absolute values of the negative portion of the signal from Frame 1, the positive values of the signal, the third component time-domain signal can be reconstructed. An FFT is taken of this signal and the even subcarriers are selected to retrieve the signal to recover the symbols for the third component.

ASCO-OFDM is designed to be more spectral efficiency than ACO-OFDM. However, it does not quite reach the spectral efficiency of DCO-OFDM or HACO-OFDM. ASCO-OFDM has a fixed spectral efficiency and like HACO-OFDM, additional components are not possible. The overall bit rate for ASCO-OFDM is given 
\begin{equation} \label{ASCO_rate}
R_{\textrm{ASCO}} = \frac{N/2-(N/4-1)}{2(N+N_{\textrm{CP}})}B\log_{2}M~bits/s
\end{equation}

\subsection{eU-OFDM}

Enhanced unipolar OFDM (eU-OFDM) is introduced in \cite{th14}. This method differs from SEE-OFDM, HACO-OFDM, and ASCO-OFDM in that it does not rely on an ACO-OFDM-like signal for any of the components at all but instead on unipolar OFDM (U-OFDM), which itself is based off DCO-OFDM. However, it is similar to SEE-OFDM in that it can use a variable amount of components and that the number of components determine the spectral efficiency of the format. In fact, an $r$ number component for both SEE-OFDM and eU-OFDM give similar spectral efficiency.

First, U-OFDM is described since eU-OFDM is just a combination of several U-OFDM signals along multiple components and frames, see Fig.~\ref{frames}-e. U-OFDM is created by modulating both even and odd subcarriers similar to DCO-OFDM, frequency-domain vector given
\begin{equation} \label{Unipolar}
  X_{\textrm{U}} = [0, X_{\textrm{0}}, X_{\textrm{1}}, ..., X_{\textrm{N/2-2}}, 0, X_{\textrm{N/2-2}}^*, ..., X_{\textrm{1}}^*, X_{\textrm{0}}^*]
\end{equation}
However, unlike DCO-OFDM, a DC-bias is not used. Instead the time-domain signal obtained from the IFFT operation, $x_{\textrm{U}}$ is divided into a positive frame and a negative frame. $x_{\textrm{U,N}}$ is equal to the negative values of $x_{\textrm{U}}$ zero clipped and $x_{\textrm{U,P}}$ is equal to the positive values of $x_{\textrm{U}}$ zero clipped and the absolute value taken of the remaining signal. Therefore, an U-OFDM signal consists of two frames: $x_{\textrm{U,N}}$ and $x_{\textrm{U,P}}$ sent consecutively.

eU-OFDM builds on this format of OFDM by using multiple U-OFDM signals to increase parallelism for a multi-component implementation similar to SEE-OFDM. Following Fig.~\ref{frames}-e, multiple components can be added together, following the same format briefly introduced above. For one, each additional component requires sending each individual U-OFDM part a factor of 2 times more than the previous component. For example, if the first component only required one of each U-OFDM part, the second component requires two, and the third component requires four, etc. Because of the repetition, the power of individual components is split among the associated parts. Therefore, each part is divided by the number of repetitions implemented to obtain proper power scaling. This scaling factor is defined as $1/\sqrt{2^{d-1}}$, where $d$ indicates the component. For example, the second component requires each of the two parts of the component to be sent twice and also each of the amplitudes to half. The repetition required in the subsequent create additional frames. Therefor, in order to complement the signal length, new frames are added to the prior components.  In summary, the more components implemented, the more frames are required before demodulation can occur. 

Like SEE-OFDM, these signals can be iteratively decoded at the receiver. However, there is a reconstruction step in between, where each negative frame is subtracted from the positive frame. The first component is decoded first. The noise from successive components do not interfere with the first component due to noise cancellations during the subtraction operation \cite{th14}. Once data from the first component is recovered, the data is remodulated accordingly to estimate the noise interference on the subsequent components.  Important to note during the subsequent components is the resuming of the parts of the same signal before subtracting the negative parts from the positive parts. Subtraction and remodulation is done iteratively until all the data is recovered. 

eU-OFDM has a similar spectral efficiency as SEE-OFDM. The spectral efficiency depends on the number of components. Here is the overall bit rate
\begin{equation} \label{eU_rate}
R_{\textrm{eU}} = \frac{N/4-1}{(N+N_{\textrm{CP}})}\sum_{d=1}^{D}\frac{1}{2^{d-1}}B\log_{2}M
\end{equation}

The main difference between SEE-OFDM and eU-OFDM is that eU-OFDM relies on the U-OFDM format and SEE-OFDM relies on the ACO-OFDM format. Both base formats have the same spectral efficiency and both enhanced formats also have the same spectral efficiency. ACO-OFDM is obtains its spectral efficiency through frequency-domain subcarrier selection whereas U-OFDM obtains its spectral efficiency through time-domain construction.

\section{Comparison amongst all techniques} \label{comp_all}

In this section, results from Monte Carlo simulations of ACO-OFDM, SEE-OFDM, HACO-OFDM, eU-OFDM, and ASCO-OFDM are presented. For each simulation, an $N=64$-length IFFT/FFT  signal is used as well as a hard-decision subtraction iterative receiver. The average electrical signal power ranges from  4 dBm to 30 dBm. At the receiver, shot and thermal noise are modeled as AWGN  with an average power of -15 dBm. Accordingly, the system SNR is in the range of 10 dB to 45 dB, which is in the acceptable reported range for indoor VLC systems.

\begin{table*}[!htb]
\centering
\small
\begin{tabular}{|c|cc|cc|cc|}
\hline
            & \multicolumn{2}{|c|}{Low Rates} & \multicolumn{2}{|c|}{Medium Rates} & \multicolumn{2}{|c|}{High Rates} \\
            & QAM/PAM   & Rate (bits/s/Hz)  & QAM/PAM    & Rate (bits/s/Hz)    & QAM/PAM   & Rate (bits/s/Hz)   \\
						\hline
ACO         & 64        & 1.50              & 128        & 1.75                & N/A*      & N/A*                \\
HACO        & 8         & 1.45              & 16         & 1.94                & N/A*      & N/A*                \\
SEE-2       & 16        & 1.50              & 32         & 1.88                & 64        & 2.25               \\
SEE-3       & 8         & 1.31              & 16         & 1.75                & 32        & 2.18               \\
SEE-3-Mix   & 8/16      & 1.50              & 16/32      & 1.94                & 32/64     & 2.38               \\
ASCO        & 16        & 1.47              & 32         & 1.84                & 64        & 2.20               \\
eU          & 16        & 1.45              & 32         & 1.82                & 64        & 2.18               \\
\hline
\end{tabular}
\caption{QAM/PAM Modulation order and bit rates used for Fig.~12. *N/A indicates that a BER of $10^{-4}$ for that modulation order and rate is not feasible when constrained to a linear regime.}
\label{modulation}
\end{table*}
\begin{figure}[!htb]
  \centering
  \includegraphics[scale=0.65]{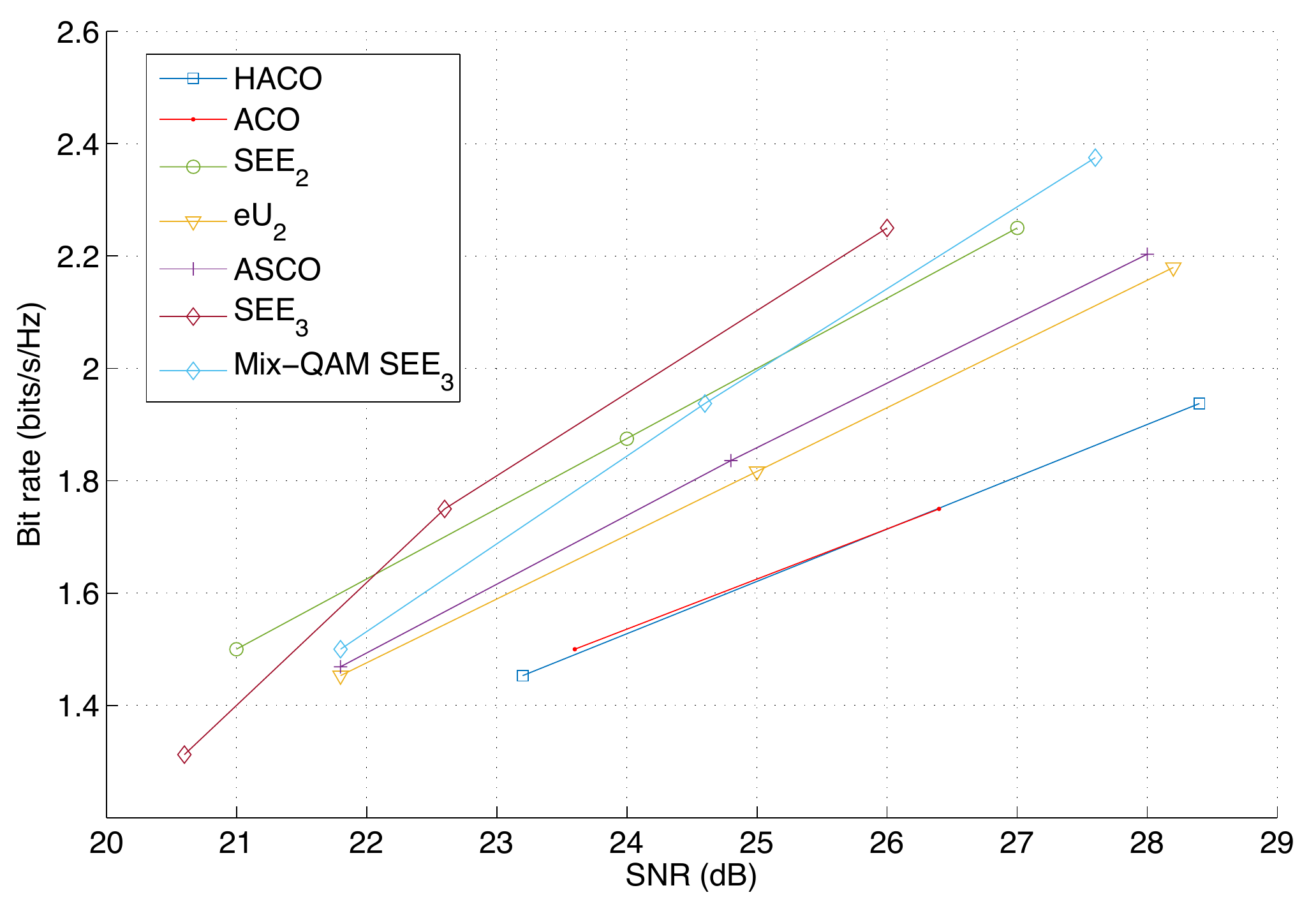}
  \caption{SNR required for given bitrate to achieve a BER of $10^{-4}$. The leftmost curve, $SEE_{3}$, has the best bit rate per SNR value for SNRs greater than 22 dB.}\label{SNRvsBitRate}
\end{figure}
For Fig.~\ref{SNRvsBitRate}, SNR required for a $10^{-4}$ BER performance plotted against the bit rate. $\textrm{SEE}_{\textrm{2}}$ and $\textrm{SEE}_{\textrm{3}}$ correspond to a 2-component and 3-component SEE-OFDM implementation. Mix-QAM-$\textrm{SEE}_{\textrm{3}}$ stands for a mixed QAM implementation where different components of the SEE-OFDM used different order of QAM modulation. eU stands for a 2-component implementation. The modulation order for the QAM/PAM and the corresponding rates can be found in Table~\ref{modulation}. The three leftmost curves in Fig.~\ref{SNRvsBitRate} are all SEE-OFDM, showing SEE-OFDM's dominance over the other techniques. Using a 2-component SEE-OFDM as a baseline, SEE-OFDM is 3 dB and 3.5 dB better than ACO-OFDM for low and high modulation orders, respectively. The two-component SEE-OFDM is also at least 1.5 dB better than ASCO-OFDM and eU-OFDM. HACO-OFDM performs similar to ACO-OFDM, despite the fact that HACO-OFDM has a better spectral efficiency than ACO-OFDM. This is because HACO-OFDM is hurt by the PAM modulation required for the second component, which requires a higher SNR than QAM modulation.
\begin{figure}[!htb]
  \centering
  \includegraphics[scale=0.65]{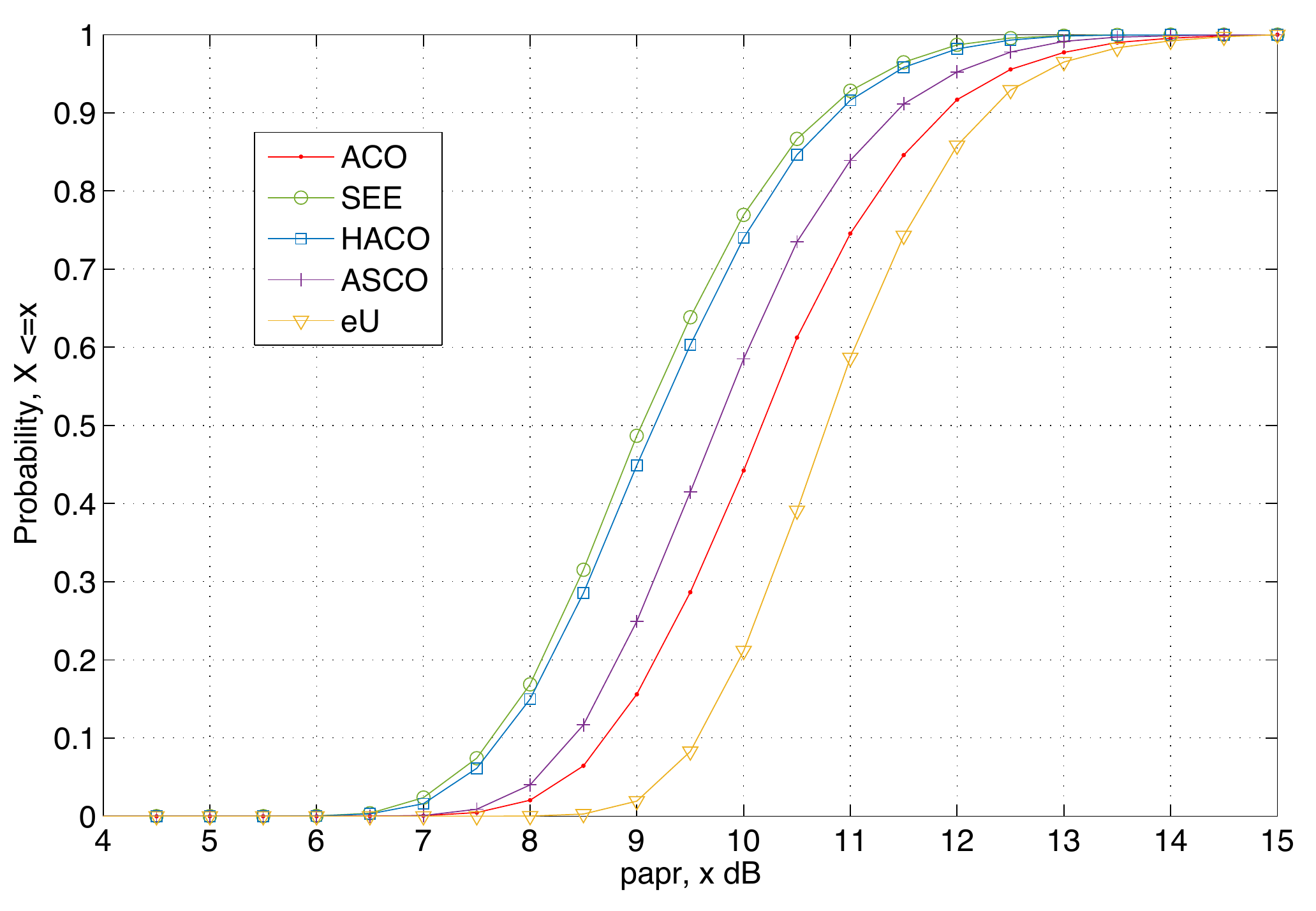}
  \caption{PAPR performance comparing combination OFDM to ACO-OFDM. Leftmost curve has the best performance.}\label{PAPR_ALL}
\end{figure}
PAPR is important when considering dynamic range. SEE-OFDM again has the better performance with a lower PAPR than the rest of the techniques, Fig.~\ref{PAPR_ALL}. Regarding PAPR, SEE-OFDM performs 1.5 dB better than eU-OFDM, 0.7 dB better than ASCO-OFDM and 1.2 dB better than ACO-OFDM.

\section{Conclusion}\label{conclusions}

SEE-OFDM is a multi-component optical OFDM technique. This allows for
different power allocation among the components. We found that when
power was equally distributed between all components and a
hard-decision iterative receiver was used, it had a 1.5 dB improvement
over other IM/DD multicarrier methods. On the receiver end,
hard-decision subtraction accounts for a 1.5 dB better performance
than reconstruction and soft-decision subtraction. Compared to other
techniques, comparing SNR for a fixed BER, SEE-OFDM is at least 1.5 dB
better than ASCO-OFDM and eU-OFDM and 3 dB better than HACO-OFDM and
ACO-OFDM at equal bit rates. SEE-OFDM also has the lowest PAPR
compared to ACO-OFDM, HACO-OFDM, ASCO-OFDM, and eU-OFDM. In addition,
SEE-OFDM is straightforward to implement. It does not require
additional frames to implement and has the capability of being
generated completely in the frequency-domain. In other words, after
the IFFT and zero-clipping, each component does not require additional
manipulation of the signal before summation. Iterative subtraction
receivers add to SEE-OFDM's easy implementation. These quantitative
and qualitative advantages of SEE-OFDM over other combination
techniques make SEE-OFDM the ideal modulation technique for future
IM/DD OWC systems.



\appendices

\section{SEE-OFDM and different length ACO-OFDM signals} \label{timefreq}
To show that replicating the smaller ACO-OFDM component in time is equivalent to modulating the even-odd subcarriers,  we have the following proof. Consider the $n^{th}$ component, modulated on the $(2k+1)2^{n}$ subcarriers:
\begin{eqnarray*}
x_n &=& \sum_{k=0}^{N/8-1} c_k e^{j2\pi (2k+1)2 n/N} \\
&=& \sum_{n=0}^{N/8-1} c_k e^{j2\pi (2k+1)n/(N/2)}
\end{eqnarray*}
This is the same as modulating an $N/2$ point ACO-OFDM signal. Note also that
\begin{eqnarray*}
x_{n+N/4}& =&\sum_{k=0}^{N/8-1} c_k e^{j2\pi 2(2k+1)(n+N/4)/N} \\
&=& \sum_{k=0}^{N/8-1} c_k e^{j2\pi 2(2k+1)n/N} e^{j2\pi (2k+1)/2} \\
&=& -x_n
\end{eqnarray*} 
Also
\begin{eqnarray*}
x_{n+N/2} &=& \sum_{k=0}^{N/8-1} c_k e^{j2\pi 2(2k+1)(n+N/2)/N} \\
&=& \sum_{k=0}^{N/8-1} c_k e^{j2\pi 2(2k+1)n/N} e^{j2\pi (2k+1)} \\
&=& x_{n}
\end{eqnarray*}
So in this case, you replicate the N/2 length ACO OFDM signal in time which is what is being done in the time-domain generation.

\section{No Interference on previous components}\label{interference}
Note that successive $x_{\textrm{SEE},p}$ signals do not
interfere with the prior
$x_{\textrm{SEE},1},\cdots,x_{\textrm{SEE},(p-1)}$ signals. The fact
that the second component, $x_{\textrm{SEE},2}$, does not interfere
with the prior SEE-OFDM component, $x_{\textrm{SEE},1}$, can be
explained as follows. Assuming that $N^{+}$ represents the set of
$n$-indices such as $x_{\textrm{SEE}}(n)>0$ and $N^{-}$ represents the
set of $n$-indices such as $x_{\textrm{SEE}}(n)<0$. The receiver
decodes the time-domain SEE-OFDM symbol $x_{\textrm{SEE}}$ by
performing the FFT operation. The FFT of the unclipped
$x_{\textrm{SEE},2}(n)$ at the odd subcarriers $(2k+1)$ of the first
component can be described as,
\begin{align}
\textrm{FFT}[x_{\textrm{SEE},2}(n)]_{(2k+1)}=\frac{1}{N}\sum_{n=0}^{N-1}x_{\textrm{SEE},2}(n) ~e^{\frac{-j2\pi (2k+1)n}{N}}\nonumber \\
= \frac{1}{N}\sum_{N^{+}}x_{\textrm{SEE},2}(n) ~e^{\frac{-j2\pi (2k+1)n}{N}}\nonumber\\
+ \frac{1}{N}\sum_{N^{-}}x_{\textrm{SEE},2}(n) ~e^{\frac{-j2\pi (2k+1)n}{N}}\nonumber\\
\label{interference_a}
\end{align}

By manipulating the right-hand side of the equation, 
\begin{align}
\textrm{FFT}[x_{\textrm{SEE},2}(n)]_{(2k+1)}=\frac{1}{N}\sum_{N^{+}}x_{\textrm{SEE},2}(n) ~e^{\frac{-j2\pi (2k+1)n}{N}}\nonumber\\
- \frac{1}{N}\sum_{N^{+}}x_{\textrm{SEE},2}(n) ~e^{\frac{-j2\pi (2k+1)(n-N/4)}{N}}\nonumber\\
\label{interference_b}
\end{align}

Therefore
\begin{align}
\textrm{FFT}[x_{\textrm{SEE},2}(n)]_{(2k+1)}=\frac{1}{N}\sum_{N^{+}}x_{\textrm{SEE},2}(n) ~e^{\frac{-j2\pi (2k+1)n}{N}}\nonumber\\
- \frac{1}{N}\sum_{N^{+}}x_{\textrm{SEE},2}(n) ~e^{\frac{-j2\pi (2k+1)(n)}{N}}e^{\frac{j\pi (2k+1)}{2}}\nonumber\\
\label{interference_c}
\end{align}

and finally,
\begin{align}
\textrm{FFT}[x_{\textrm{SEE},2}(n)]_{(2k+1)}=\frac{1}{N}\sum_{N^{+}}x_{\textrm{SEE},2}(n) ~e^{\frac{-j2\pi (2k+1)n}{N}}\nonumber\\
\times(1-j^{(2k+1)})\nonumber\\
\label{interference_d}
\end{align}

The left hand side in equation (\ref{interference_d}) is the FFT of the unclipped signal and
is equal to zero by construction. The right hand side is the the FFT of the clipped version of the 
signal multiplied by $(1-j^{2k+1})$. Since the term $(1-j^{2k+1})$ is not equal to zero, the FFT of the 
clipped signal must equal to zero. Therefore the $2k+1$, odd, subcarriers 
must also equal to zero, and there is no interference.
Although the latter components do not interfere with the prior
components, the prior components do interfere with the additional
components. For example for a 2-component SEE-OFDM signal, the first
component has no interference from the second component. However, the second component has
interference from the first.  This interference can be addressed at the receiver
end by the reconstruction step or successive
interference cancellation.

\section*{Acknowledgement}
This work is supported by the National Science Foundation under Grant No. EEC-0812056.

\bibliographystyle{IEEEtran}
\bibliography{references}

\end{document}